\documentclass[twocolumn]{aastex63}

\usepackage{graphicx}	
\usepackage{color}
\usepackage{amsmath}	% Advanced maths commands
\usepackage{amssymb}	% Extra maths symbols
\usepackage{mathrsfs}
\usepackage{mathrsfs}

\newcommand{\msun}{M_\odot}

\newcommand{\zsun}{Z_\odot}
\newcommand{\lsun}{L_\odot}

\newcommand{\cc}{{\rm cm}^{-3}}

\newcommand{\msunyr}{M_\odot~{\rm yr}^{-1}}

\newcommand{\kpc}{{\rm kpc}}

\newcommand{\pc}{{\rm pc}}
\newcommand{\mum}{\mu {\rm m}}
\newcommand{\kms}{{\rm km~s}^{-1}}

\newcommand{\K}{{\rm K}}
\newcommand{\beq}{\begin{equation}}
\newcommand{\eeq}{\end{equation}}

\shorttitle{Spectra of Early Seed BHs}
\shortauthors{Inayoshi et al.}

\begin{document}

\title{The Age of Discovery with the James Webb: \\Excavating the Spectral Signatures of the First Massive Black Holes}

\correspondingauthor{Kohei Inayoshi}
\email{inayoshi@pku.edu.cn}

\author[0000-0001-9840-4959]{Kohei Inayoshi}
\affiliation{Kavli Institute for Astronomy and Astrophysics, Peking University, Beijing 100871, China}
\author[0000-0003-2984-6803]{Masafusa Onoue}
\altaffiliation{Kavli Astrophysics Fellow}
\affiliation{Kavli Institute for Astronomy and Astrophysics, Peking University, Beijing 100871, China}
\affiliation{Kavli Institute for the Physics and Mathematics of the Universe (Kavli IPMU, WPI), The University of Tokyo, Chiba 277-8583, Japan}
\author[0000-0001-6958-7856]{Yuma Sugahara}
\affiliation{Waseda Research Institute for Science and Engineering, Faculty of Science and Engineering, Waseda University, 3-4-1, Okubo, Shinjuku, Tokyo 169-8555, Japan}
\affiliation{National Astronomical Observatory of Japan, 2-21-1 Osawa, Mitaka, Tokyo 181-8588, Japan}
\author[0000-0002-7779-8677]{Akio K. Inoue}
\affiliation{Waseda Research Institute for Science and Engineering, Faculty of Science and Engineering, Waseda University, 3-4-1, Okubo, Shinjuku, Tokyo 169-8555, Japan}
\affiliation{Department of Physics, School of Advanced Science and Engineering, Faculty of Science and Engineering, Waseda University, 3-4-1, Okubo, Shinjuku, Tokyo
169-8555, Japan}
\author[0000-0001-6947-5846]{Luis C. Ho}
\affiliation{Kavli Institute for Astronomy and Astrophysics, Peking University, Beijing 100871, China}
\affiliation{Department of Astronomy, School of Physics, Peking University, Beijing 100871, China}

\begin{abstract}
The James Webb Space Telescope (JWST) will open a new window of the most distant universe
and unveil the early growth of supermassive black holes (BHs) in the first galaxies.
In preparation for deep JWST imaging surveys, it is crucial to understand the color selection of high-redshift 
accreting seed BHs.
We model the spectral energy distribution of super-Eddington accreting BHs with millions of solar masses in metal-poor galaxies at $z\ga 8$,
applying post-process line transfer calculations to radiation hydrodynamical simulation results.
Ten kilosecond exposures with the NIRCam and MIRI broad-band filters are sufficient to detect the radiation flux from the seed 
BHs with bolometric luminosities of $L_{\rm bol}\simeq 10^{45}~{\rm erg~s}^{-1}$.
While the continuum colors are similar to those of typical low-$z$ quasars, 
strong H$\alpha$ line emission with a rest-frame equivalent width ${\rm EW}_{\rm rest}\simeq 1300~{\rm \AA}$ is so prominent 
that the line flux affects the broad-band colors significantly. 
The unique colors, for instance F356W$-$F560W $\ga  1$ at $7<z<8$ and F444W$-$F770W $\ga  1$ at $9<z<12$, provide robust criteria
for photometric selection of the rapidly growing seed BHs.
Moreover, NIRSpec observations of low-ionization emission lines can test whether the BH is fed via a dense accretion disk 
at super-Eddington rates.
\end{abstract}


\section{Introduction}
The existence of massive black holes (BHs) with $M_\bullet \gtrsim 10^8~\msun$ observed when the universe was younger than one billion years 
strongly constrains their origin and formation pathway \citep{Wu_2015,Matsuoka_2018,Banados_2018,Wang_2021}.
Their quick assembly has been attributed to mechanisms such as formation of heavy seeds as massive as $\sim 10^{3-5}~\msun$
and rapid gas accretion onto their seeds \citep[e.g.,][]{Haiman_2013,Inayoshi_ARAA_2020,Volonteri_2021}.

One of the most interesting questions is whether or not BH seeding and growth models are distinguishable 
using current and future observation facilities. 
The most direct answers will be obtained through the detection of BHs with lower masses of $M_\bullet \lesssim 10^{7-8}~\msun$ at $z>7$
(note that the lowest BH mass measured at $z>6$ with the \ion{Mg}{2} single-epoch method is $3.8^{+1.0}_{-1.8}\times 10^7~\msun$; 
see \citealt{Onoue_2019}).
Recent radiation-hydrodynamic (RHD) simulations suggest that seed BHs formed in massive galaxies that end up 
in high-$z$ quasar hosts can experience multiple accretion bursts at super-Eddington rates
and grow in mass \citep{Inayoshi_2022}.
This process naturally brings the BH mass above the BH-galaxy mass correlation seen in the local universe \citep[e.g.,][]{Kormendy_Ho_2013}.
Moreover, the existence of such overmassive BHs provides us with a unique opportunity to detect seed BHs in the very early universe 
with upcoming deep near-infrared observations, such as the James Webb Space Telescope (JWST) and Nancy Grace Roman Space Telescope.  

In this {\it Letter}, we model the spectral energy distribution (SED) of a rapidly growing BH in the nucleus of a protogalaxy,
which is approximately $10^{2-3}$ times fainter than the most luminous quasars known at $z>6$, but is bright enough 
to be detected with the upcoming JWST imaging surveys thanks to its high accretion rate. 
We show that multi-band photometry with NIRCam and MIRI enables us to robustly select this extremely young BH 
when it appears at $z\sim 7-12$. 
Moreover, follow-up spectroscopic observations of the candidates identified by this method will be useful to diagnose the stage of the BH 
growth phase and the role of BH feedback.
Throughout this {\it Letter}, we assume a $\Lambda$ cold dark matter cosmology consistent with the latest constraints from Planck
\citep{Planck_2020}; $h=67.66$, $\Omega_{\rm m}=0.3111$, and $\Omega_\Lambda =0.6889$.

\section{SED modeling of high-$z$ seed BHs}

To model SEDs of a fast-growing BH, we make use of {\tt CLOUDY} \citep[C17;][]{Ferland_2017} to apply post-process line transfer 
calculations to the results of two-dimensional RHD simulations where the nuclear scale at $0.1~\pc \leq r \leq 100~\pc$ is well-resolved \citep{Inayoshi_2022}.
We consider the model with a high star formation efficiency in the bulge of a metal-poor galaxy ($Z=0.01~\zsun$ for gas and stars), 
where a seed BH with an initial mass of $M_{\rm \bullet, ini}=10^5~\msun$
is fed via a massive gaseous disk efficiently at rates of $\sim 1~\msunyr$ exceeding the Eddington value, 
and its mass increases nearly tenfold within $1$ Myr.
The elapsed time of $t=3$ Myr since the simulation begins (corresponding to $\simeq 1$ Myr after the last accretion burst)
is adopted because the accretion flow settles down to a steady state.
This case yields the highest bolometric luminosity of $L_{\rm bol}\simeq 10^{45}~{\rm erg~s}^{-1}$
compared to the other models where the BH accretes at super-Eddington rates but its mass does not reach $10^6~\msun$ within 
the computational time, which is not heavy enough to produce the same level of luminosity \citep[see more details in][]{Inayoshi_2022}.
In this work, we do not specify the BH seeding mechanisms\footnote{
We do not exclude a scenario that a less massive seed with $M_\bullet \ll 10^5~\msun$ at birth 
\citep[e.g.,][]{Sassano_2021} is falling into the galactic center through halo mergers 
and its rapid growth to $\sim 10^5~\msun$ via mass accretion is viable \citep{Ryu_2016b}.}
but focus on the observational signature at the early growth stage when the BH mass approaches $M_\bullet \simeq 10^6~\msun$.

%%%%%%%%
%	Fig. 1     %
%%%%%%%%
\begin{figure}
\begin{center}
{\includegraphics[width=87mm]{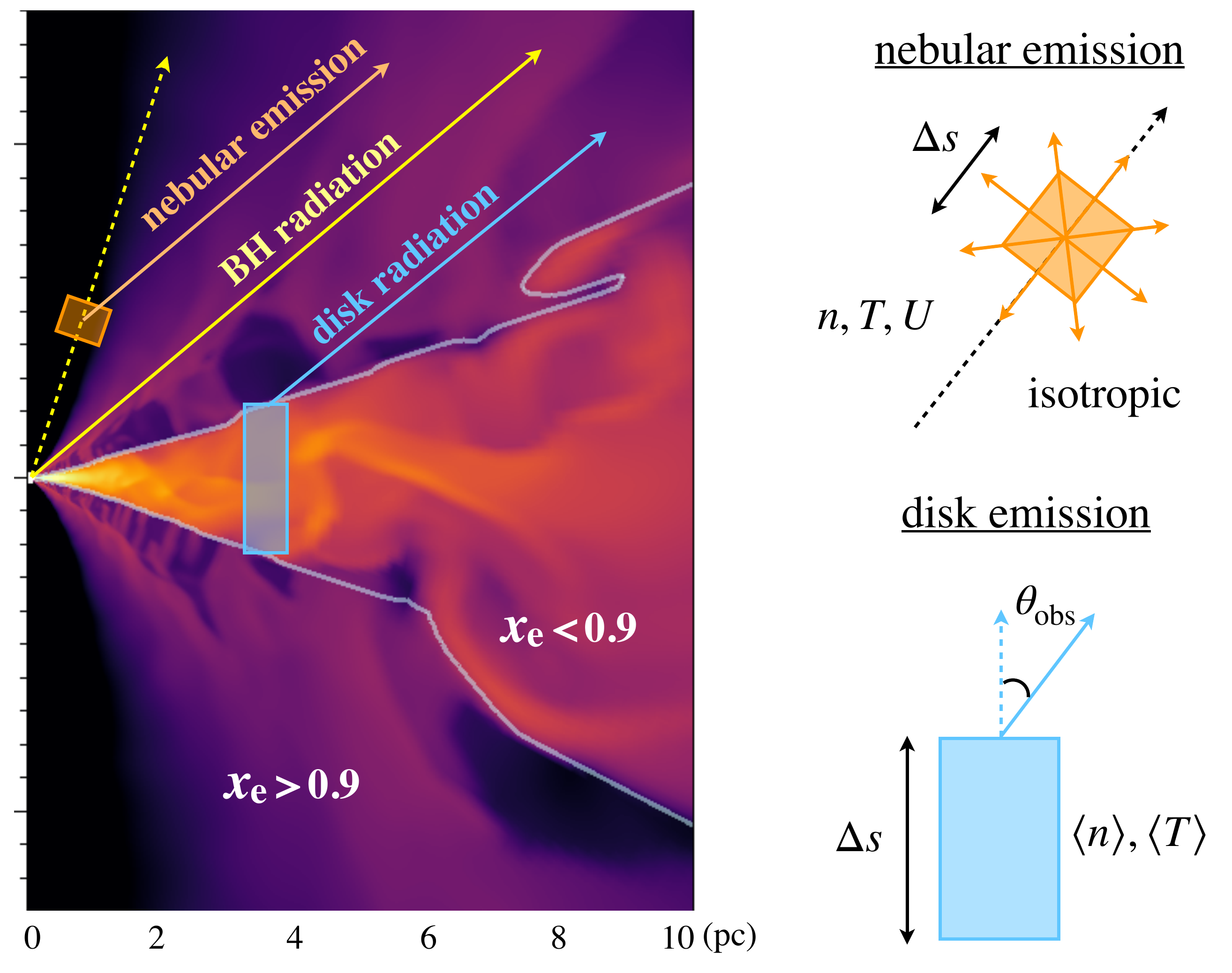}}
\caption{Density structure of the accretion flow onto a seed BH and a schematic picture describing our SED modeling,
which includes three components: (1) the radiation flux produced from the unresolved nuclear disk of the BH,
(2) nebular emission lines emitted from irradiated gas parcels, and (3) radiation from the dense accreting disk
in the RHD simulation domain.
The physical quantities of $n$, $T$, $U$, $x_{\rm e}$, $\Delta s$, and $\theta_{\rm obs}$ are 
the number density of hydrogen nuclei, gas temperature, ionization parameter, electron fraction, thickness of gas parcels,
and the viewing angle, respectively.
The quantities with brackets $\langle \cdot \rangle$ are the mass-weighted values along the vertical direction (see more details in the text).
The boundary between the nebula and disk is defined by $x_{\rm e}=0.9$ (thin curve).
}
\label{fig:cartoon}
\end{center}
\end{figure}

Figure~\ref{fig:cartoon} shows the structure of the accretion flow in the RHD simulation and 
three components of radiation considered in our SED model.
The direct component of the flux associated with BH accretion is the primary source of radiation.
The radiation luminosity injected from the unresolved central region to the computational domain
is assumed to follow a broken-power law SED as
\begin{align}
L_{\bullet, \nu} = \begin{cases}
L_0 \left( \dfrac{\nu}{\nu_0}\right)^{-0.6} & (\nu_{\rm min} \leq \nu < \nu_0), \vspace{10pt}\\
L_0 \left( \dfrac{\nu}{\nu_0}\right)^{-1.5} & (\nu_0 \leq \nu \leq \nu_{\rm max}),
\end{cases}
\label{eq:specin}
\end{align}
\citep[e.g.,][]{Lusso_2015}, where $h\nu_0 = 10~{\rm eV}$, $h\nu_{\rm min} = 1~{\rm eV}$, and 
$h\nu_{\rm max} = 1~{\rm keV}$.
The normalization of the luminosity is given by an analytical model for the standard and slim disk solutions \citep{Watarai_2000}.
Photoheating of the accretion flow is dominated by bound-free absorption of hydrogen and helium 
in the UV and soft X-ray bands but X-ray heating of heavy elements is subdominant in the low-metallicity environments.
Note that the hardness of the incident radiation flux hardly affects the conditions for the onset of super-Eddington accretion
\citep[e.g.,][]{Takeo_2019}.
Furthermore, we assume isotropic and anisotropic radiation fields depending on the bolometric luminosity emitted from the accreting BH
\citep[see more details in][]{Inayoshi_2022}.
This model injects non-zero radiation flux to the equatorial region ($\theta \simeq \pi/2$) up to the Eddington flux
and affects the thermal properties of the accretion disk.
The reprocessed component of radiation is considered to be the disk emission (see below).
The luminosity in excess of the Eddington value is injected anisotropically 
so that the flux is collimated to the poles as $\propto (\cos \theta)^4$.

The simulation domain is separated into nebular gas irradiated by the luminous accreting BH 
and a dense disk that feeds the central BH.
Here, we define the boundary of the two regions as that where the electron fraction is $x_{\rm e}=0.9$, which gives a clear separation between them 
because the mean free path at the ionization front is shorter than the single cell size.
The nebular gas reprocesses emission lines associated with recombination of ionized atoms.
The ionization parameter $U$ for each cell is calculated with the $r$-component of the radiation flux (attenuation by atoms and dust grains is included).
The gas density and temperature are taken from the RHD simulation, and the slab thickness $\Delta s$ is set to the cell size  
along the radial direction.
Thus, the nebular line luminosity is calculated as
\begin{equation}
L_{\rm line} =\int_{\mathcal{N}}  4\pi j_{\rm line} dV ,
\end{equation}
where $4\pi j_{\rm line}$ is the frequency-integrated line emissivity and isotropic emission is considered.
The volume integration is conducted through the nebular region where photoionization of gas dominates ($x_{\rm e}>0.9$).
The nebular emission is mainly produced at larger radii, although the computation domain
of the RHD simulation is limited within $100~\pc$.
To evaluate the contribution from $r>100~\pc$, we extrapolate the gas properties and strength of ionizing radiation to
the virial radius of the host dark matter halo $r_{\rm vir}\simeq 1.6~\kpc$, assuming the gas density to follow 
$n(r)=n_0(r/100~\pc)^{-\alpha}$.
Here, the characteristic density and power-law index are set to $n_0=30~\cc$ and $\alpha=2$, consistent with the RHD simulation
and cosmological simulations for galaxy formation \citep[e.g.,][]{Regan_2014}.
With the extrapolated density profile, the ionization front expands without disturbing the density structure \citep{Franco_1990}.
The nebular emission from the outer region contributes to $70~\%$ of Ly$\alpha$ and $50~\%$ of \ion{He}{2} line flux 
on the total SED, and the other emission lines are produced from the dense accretion disk within $1~\pc$.
Note that the \ion{He}{2} line intensity powered by the accreting BH with a hard spectrum (see Eq.~\ref{eq:specin}) is maintained unless 
the incident spectrum is as soft as $\beta \gtrsim 2$, where $L_{\bullet,\nu} \propto \nu ^{-\beta}$.

%%%%%%%%
%	Fig. 2     %
%%%%%%%%
\begin{figure*}
\begin{center}
{\includegraphics[width=140mm]{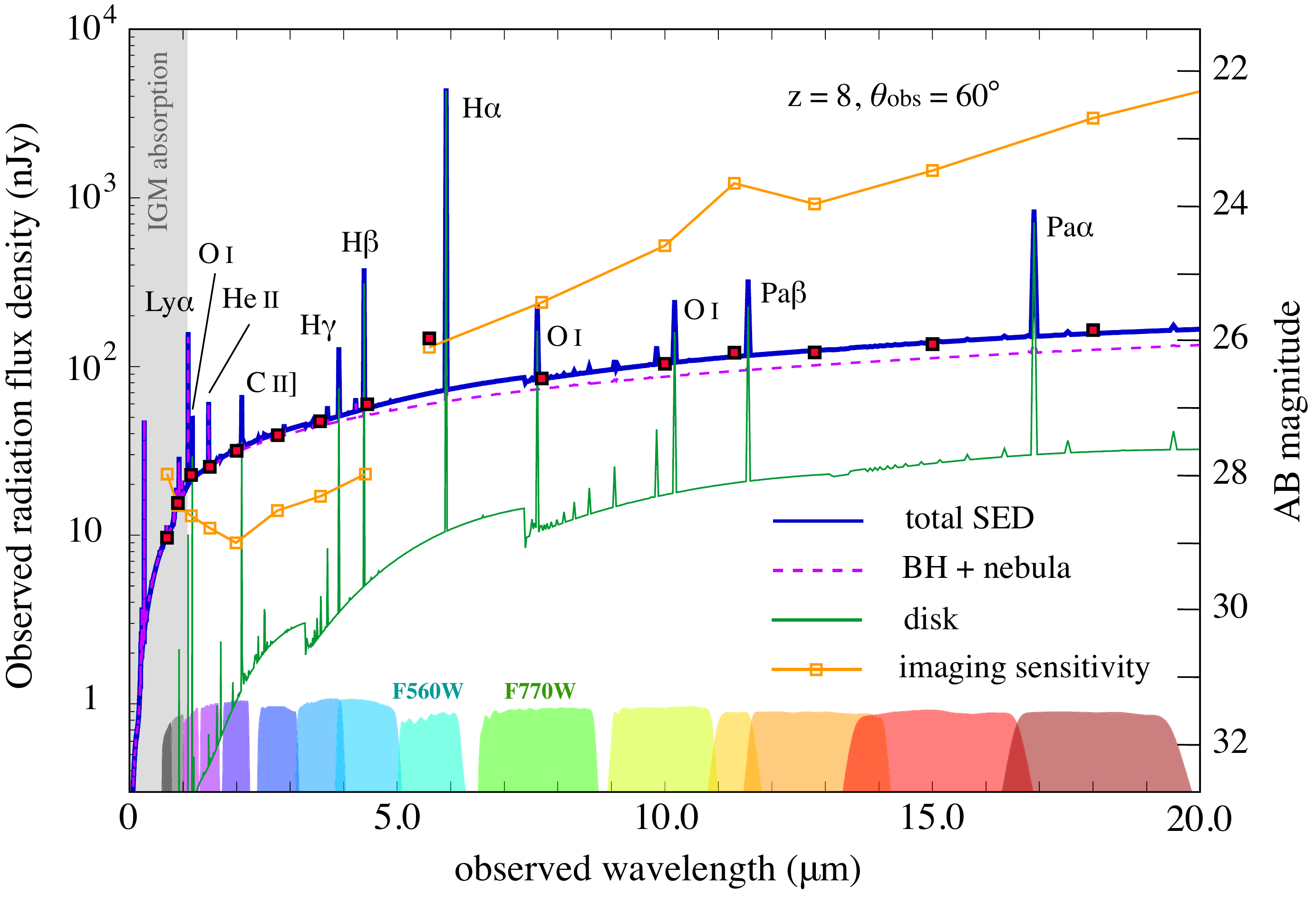}}
\caption{Spectral energy distribution of the growing seed BH with $M_\bullet \simeq 10^6~\msun$ at $z=8$:
the total SED (blue), the radiation flux from the nuclear BH with nebular emission lines (magenta), 
and the emission from the dense accretion disk at $r\sim 0.1-1~\pc$ (green). 
The viewing angle is set to $\theta_{\rm obs}=60^\circ$.
The imaging sensitivity curves with S/N=10 of JWST's NIRCam (0.6--5 $\mu$m) and MIRI (5--20 $\mu$m) in a 10 ks exposure time 
are overlaid (the open square symbol indicates the effective wavelength of each filter), 
along with the transmission curve of each filter at the bottom (arbitrary units).
The filled square symbols present the filter-convolved flux density at each filter.
The continuum radiation flux with several prominent lines with the rest-frame equivalent widths of EW$_{\rm rest}>7~{\rm \AA}$
is observable with the NIRCam broad-band filters except for F070W and the MIRI F560W filter.
Intergalactic medium (IGM) absorption is not explicitly included, but becomes important in the shaded region at 
$\lambda_{\rm obs}<1.09~\mum$.
}
\label{fig:SED1}
\end{center}
\end{figure*}

The disk region is dense enough to absorb the incident BH radiation flux and to reprocess the energy as lower-energy photons.
To calculate this component of radiation, we divide the disk into annuli and conduct radiation transfer calculations
along the vertical direction of each annulus by neglecting the radiation energy transport between annuli and 
incident radiation from the disk surface (i.e., the ionization parameter is set to $U=0$).
For each annulus, we calculate the mass-weighted averaged gas density $\langle n\rangle$ 
and temperature $\langle T\rangle$ along the vertical direction and use them in the CLOUDY calculation as constant (height independent) values.
The thickness is calculated as $\Delta s  \equiv N_{\rm e}/\langle n\rangle$, where $N_{\rm e}$ is the electron column density obtained from the RHD simulation.
We note that the CLOUDY calculation finds its own equilibrium, and thus the electron column density estimated with the physical thickness $2H(>\Delta s)$
tends to be at most 10 times higher
than the value $N_{\rm e}$ obtained from the RHD simulation that follows the dynamical, non-equilibrium nature of the accretion flow.
In fact, for a typical case in the disk with $\langle n \rangle = 10^9~\cc$ and $\langle T \rangle = 10^4~\K$,
the line fluxes scale with the thickness (namely, the H$\alpha$ and H$\beta$ fluxes increase as $\propto \Delta s^{1/2}$), 
while the continuum flux does not change.
Therefore, our treatment with a smaller thickness yielding $N_{\rm e}$ consistent with the RHD simulation 
gives a conservative prediction for strength of lines and their detectability.

In the post-process calculations, the radiation flux produced from the disk does not always match with the flux injected from the accreting BH 
because the cooling timescale in the disk is significantly shorter than the dynamical timescale.
To ensure energy conservation, we renormalize the disk luminosity obtained by integrating over the entire SED 
with the luminosity injected toward the disk region,
\begin{align}
L_{\rm \bullet,disk} 
&= 2.6\times 10^{43}~{\rm erg~s}^{-1}\left(\frac{M_\bullet}{10^6~\msun}\right)
\left(\frac{\sin \Theta}{0.2}\right),
\end{align} 
where $\Theta \equiv \tan^{-1} (H/R)$ estimated at the inner-most radius.
We here adopt $L_{\rm \bullet,disk}=2.3\times 10^{43}~{\rm erg~s}^{-1}$ ($M_\bullet =1.45\times 10^6~\msun$ and $ \Theta = 7^\circ$
taken from the RHD simulation).
Note that the incident BH radiation flux and reprocessed emission lines are obscured by dust in the dense disk 
for observers from nearly edge-on directions, covering $\lesssim 15~\%$ of fast-growing seed BHs.

In our SED model, we adopt the frequency mesh at the code default, $\Delta \lambda / \lambda \simeq 3.33\times 10^{-3}$
(the corresponding line width is $\simeq 1000~\kms$), so that an integral over the spectrum is consistent with the total line flux.
Therefore, the peak line fluxes are underestimated compared to those of its real spectrum with typical line width of 
$\sim 200-300~\kms$ (the order of the Keplerian velocity in the dense disk at the inner-most radii).

\vspace{1mm}
\section{Results}
\subsection{Spectral energy distribution}\label{sec:SED}

% SED
Figure~\ref{fig:SED1} shows the SED of the growing seed BH with $M_\bullet \simeq 10^6~\msun$ at $z=8$, where 
the viewing angle from the polar direction is set to $\theta_{\rm obs}=60^\circ$.
Each curve represents the radiation flux from the nuclear BH with nebular emission lines (magenta), 
the emission from the dense accretion disk at $r\sim 0.1-1~\pc$ (green), and the sum of the two components (blue), respectively.
The 10 kilosecond (ks) imaging sensitivity curves of JWST's NIRCam and MIRI\footnote{https://www.stsci.edu/jwst/instrumentation} 
are shown with the orange line, along with the transmission curve of each filter at the bottom (arbitrary units).
The continuum radiation flux is as high as $20-40$ nJy at $2\la \lambda_{\rm obs}/\mu{\rm m} \la 5$, which is detectable with the NIRCam.
There are several prominent lines powered by the central accreting BH (see Table~\ref{tab:EW}).
The Balmer lines are extremely strong: ${\rm EW}_{\rm rest}^{\rm H\alpha}\simeq 1320~{\rm \AA}$
and ${\rm EW}_{\rm rest}^{\rm H\beta}\simeq 94.9~{\rm \AA}$, in contrast to the composite spectrum of low-$z$ quasars, which show ${\rm EW}_{\rm rest}^{\rm H\alpha}\simeq 195~{\rm \AA}$
and ${\rm EW}_{\rm rest}^{\rm H\beta}\simeq 46~{\rm \AA}$ \citep{VandenBerk_2001}.
As a result, the filter-convolved F560W flux density ($f_\nu=150$ nJy; filled square) is twice higher than the continuum level 
at the same wavelength range and thus is well above the S/N=10 detection limit in a 10 ks exposure ($f_\nu=130$ nJy).
The strongest H$\alpha$ at $\lambda_{\rm obs}\simeq 5.9~\mum$ in the MIRI's F560W filter can be used for photometric selection of rapidly accreting seed BHs (see Sec.~\ref{sec:color}).

%%%%%%%%
%	Table     %
%%%%%%%%
\begin{deluxetable*}{lcccccccc}
\tablenum{1}
\tablecaption{Line properties of the SED shown in Figure~\ref{fig:SED1}}
\tablewidth{0pt}
\tablehead{
\colhead{Lines} & \colhead{Ly$\alpha$} & \colhead{\ion{O}{1}} & \colhead{\ion{He}{2}} & \colhead{\ion{C}{2}]} 
&\colhead{H$\gamma$} & \colhead{H$\beta$} & \colhead{H$\alpha$}\\
\colhead{$\lambda_{\rm rest}$ (\AA)/ $\lambda_{\rm obs}$ ($\mum$)} & \colhead{1215/1.09} & \colhead{1304/1.17} & \colhead{1640/1.48} 
& \colhead{2326/2.09} & \colhead{4340/3.91} & \colhead{4861/4.37} & \colhead{6563/5.91}
}
%\decimalcolnumbers
\startdata
%$\lambda_{\rm rest}$ (\AA) &1215 & 1304 & 1640 & 2326 & 4340 & 4861 & 6563 \\
EW$_{\rm rest}$ (\AA) & 28.5  & 11.7 & 7.77  & 9.79 & 22.7 & 94.9 & 1320  \\
%\hline
Line flux & 14.1 & 8.27 & 4.01 &5.03 & 3.20 & 8.41 & 72.5  \\
S/N (PRISM) & 4.45 & 3.97 & 3.35 & 4.31 & 6.00 & 12.48  & --\\
S/N ($R\sim 1,000$) & 5.86 & 4.37 & 2.43 & 4.44 & 5.78 & 12.17 & --
\enddata
\tablecomments{
The equivalent width is calculated in the rest frame. 
The line flux is in units of $10^{-19}$ erg s$^{-1}$ cm$^{-2}$.
The signal-to-noise ratios (S/N) for individual lines of the SED ($z=8$)
in a 30 ks exposure time of NIRSpec Fixed-Slit observations are calculated with the JWST Exposure Time Calculator Tool \dag,
where a low-resolution (PRISM, $R\sim 100$) and a medium-resolution ($R\sim 1,000$) mode are considered, respectively.
The line width of each line is assumed to be $300~\kms$ for the S/N calculation.
The Ly$\alpha$ damping effect by IGM absorption is not included here, but would be significant for the sources at $z\ga 8$.\\
\dag: https://jwst.etc.stsci.edu/}
\vspace{-5mm}
\label{tab:EW}
\end{deluxetable*}

The Ly$\alpha$ line, which is the most prominent emission line in the rest-frame UV spectrum of quasars, 
is not striking even for this luminous object 
because Ly$\alpha$ photons undergo a large number of scattering events within the dense accretion disk, where electron collisions
populate hydrogen atoms from $n=2$ to higher levels ($n\geq 3$) before Ly$\alpha$ photons escape \citep{Kwan_Krolik_1981}.
Therefore, this process enhances the Balmer and higher series lines, as shown in Figure~\ref{fig:SED1}.
We also note that the H$\alpha$/H$\beta$ flux ratio is 3 times higher than the ratio observed in
diffuse nebulae (H$\alpha$/H$\beta \sim 2.86$ at $T=10^4~\K$) because the production of those lines in the dense disk
is dominated by the collisional excitation of hydrogen \citep[e.g.,][]{Wills_1985}.
The emission lines of neutral oxygen (\ion{O}{1}\,$\lambda\lambda 1304, 8446, 11287$) and singly ionized carbon 
(\ion{C}{2}]\,$\lambda$2326) are effectively excited in the dense gaseous disk at $0.1~\pc \la r \la 1~\pc$.
The production of the three \ion{O}{1} lines is a result of Ly$\beta$ fluorescence
that occurs when a population in $n=3$ of hydrogen is 
built up by collisional excitation and thus tightly correlates to the enhancement of Balmer lines.
The \ion{C}{2}] line is collisionally excited in the same dense region and optically thin.
These low-ionization lines support the existence of a dense disk feeding the BH at a high rate.
Those dynamical and radiative processes can be investigated with the RHD simulation that resolves the nuclear scale, 
which is not sufficiently resolved in large-scale cosmological simulations due to computational limitations.

The forbidden lines of [\ion{C}{2}] 158 $\mum$, [\ion{O}{3}] 88 $\mum$, and [\ion{O}{1}] 63 $\mum$ and 145 $\mum$
are depopulated by collisions in the dense nucleus. However, those lines would be produced from star-forming regions at larger scales. 
In our model, for instance, the luminosity of [\ion{C}{2}] emission line associated with star formation is estimated as $\sim 10^{7.7}~\lsun$ 
with an empirical relation obtained from \cite{Romano_2022}.
This level of luminosity would be marginally detectable with the Atacama Large Millimeter/submillimeter Array
for the sources at $z\sim 5$, and thus requires gravitational lensing magnification beyond $z\sim 6$ \cite[e.g.,][]{Fujimoto_2021}.

We note that high-ionization metal lines are not produced from the accreting gas onto BH seeds
in metal-poor protogalaxies ($Z=0.01~\zsun$), unlike metal-enriched quasars at lower redshifts.
Adopting the same conditions, narrow metal lines such as \ion{C}{4}\,$\lambda 1549$ and [\ion{O}{3}]\,$\lambda 5007$, usually seen in low-$z$ quasars,
appear in the SED as the metallicity increases to $Z\simeq \zsun$.
Moreover, this SED model does not take into account broad-line components that could be produced in the inner region at $r< 0.1$ pc, 
which is unresolved in our simulations.
Despite the lack of dedicated calculations, we expect a negligible contribution of possible broad lines to the broad-band colors because of the low metallicity of the accreting gas.

%%%%%%%%
%	Fig. 3     %
%%%%%%%%
\begin{figure*}
\begin{center}
{\includegraphics[width=180mm]{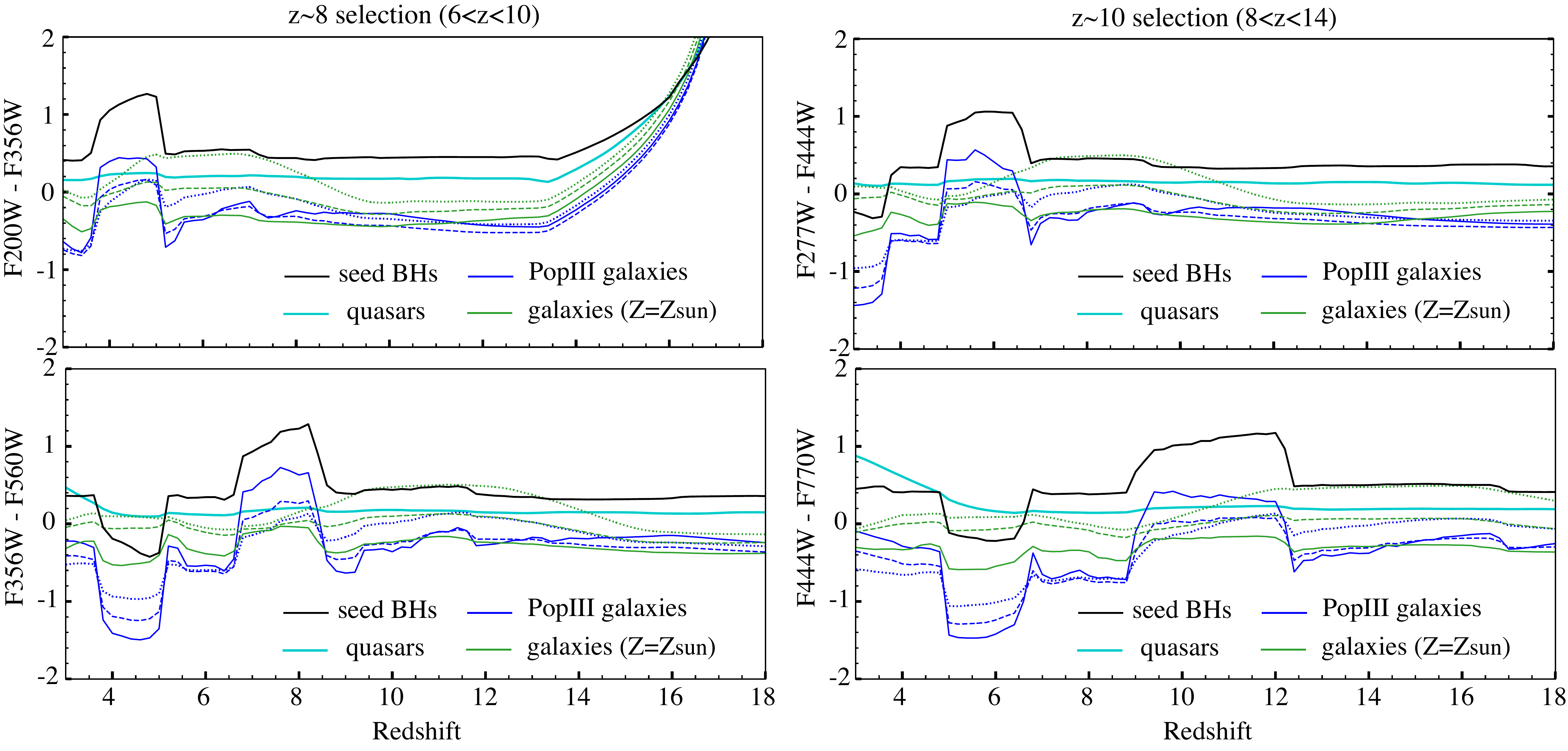}}
\caption{Redshift evolution of the broad-band colors used for selecting high-$z$ seed BHs (black)
at $z\sim 8$ (left) and $z\sim 10$ (right).
For comparison, we also plot those for other three types of high-$z$ sources: 
quasars (cyan), Population III galaxies (blue), and star-forming galaxies with $Z=\zsun$ (green)
with stellar age of $10$ (solid), $100$ (dashed), and $500$ (dotted) Myr.
The quasar SED model assumes a power-law continuum with a UV slope of $\alpha_\lambda \equiv d\ln f_\lambda/d\ln \lambda = -1.70$
\citep{Selsing_2016} and various broad emission lines, the strength of which is based on a low-$z$ measurement by \citet{VandenBerk_2001}.
The galaxy SEDs are from \citet{Inoue_2011}.
In those four panels, the broad-band colors of seed BHs are generally red compared to those of galaxies that have bluer UV continua.
The colors sharply increase (or decrease) when the prominent H$\alpha$ emission enters the longer-wavelength (shorter-wavelength) filters.
The combination of the red continuum and extremely strong H$\alpha$ makes the F356W$-$F560W and F444W$-$F770W colors of the growing seed BHs as large as $\gtrsim1$.
Note that the sharp rise of F200W $-$ F356W color at $z>14$ is due to IGM absorption.
}
\label{fig:color}
\vspace{5mm}
\end{center}
\end{figure*}

Thus far, we focus on fast-growing seed BHs at $z=8$.
Here, we briefly mention the detectability of such luminous objects at higher redshifts.
At $9<z<12$, the H$\alpha$ line enters MIRI's F770W filter and enhances the filter-convolved F770W flux density from the continuum level, as in the case at $z=8$.
For instance, the convolved flux density at $z=10$ ($f_\nu=90$ nJy) is comparable to the S/N $\ga 5$ detection limit of MIRI's F770W filter in a 20 ks exposure.
At higher redshifts, the convolved flux density in the longer-wavelength filter does not reach the detection limit in a reasonable exposure time, unless we observe the seed BHs in a face-on view ($\theta_{\rm obs}<60^\circ$).

The detection number of such accreting seed BHs is expected to be $\sim O(1)$
within ten JWST/NIRCam fields of views \citep{Inayoshi_2022}.
Therefore, when more than ten NIRCam and MIRI images are taken with sufficiently high sensitivities
and those observed regions are significantly overlapped,
there is a reasonable chance of finding at least one seed BH by upcoming JWST observations.

\vspace{1mm}
\subsection{Photometric color selection}\label{sec:color}

% z vs. color
We now discuss how the growing seed BHs can be selected with JWST broad-band imaging observations.
Here we consider two redshift ranges, namely $z\sim8$ and $z\sim10$.
Figure~\ref{fig:color} shows how the broad-band colors of the growing seed BHs (black) change as a function of redshift.
We chose two combinations of NIRCam and MIRI filters for each redshift range:
F200W vs F356W and F356W vs F560W for $z\sim8$, and 
F277W vs F444W and F444W vs F770W for $z\sim10$.
In each column, the filter choice is determined so that the continuum flux is red in the first color (top)
and the contribution from emission lines enters the filter with the longest wavelength (bottom)
at $z\sim 8$ and $z\sim 10$, respectively.
For comparison, we also plot the colors of quasars (cyan) and star-forming galaxies with $Z=0$ 
(Population III galaxies; blue) and $Z=\zsun$ (green), respectively. 
We adopt SED models for Population III galaxies and star-forming galaxies with stellar age of 
10, 100, and 500 Myr \citep{Inoue_2011}.
The quasar SED model is adopted from the low-$z$ composite spectrum \citep{VandenBerk_2001}.
Figure~\ref{fig:color} demonstrates how strongly H$\alpha$ emission impacts the observed infrared colors in specific redshift ranges.
Since the SED of accreting seed BHs shows prominent H$\alpha$ line emission, the second color in each column (bottom panels)
becomes as red as F356W$-$F560W $>1.0$ at $7<z<8$, and F444W$-$F770W $>1.0$ at $9< z < 12$, 
where the H$\alpha$ line enters the longer-wavelength filters.
The other three types of high-$z$ sources show similar evolutionary trends in those colors, 
but the high-$z$ seed BHs show the reddest ones due to 
the combination of flat continuum and H$\alpha$ emission\footnote{
Recently, \cite{Stefanon_2022} reported an extremely high value of ${\rm EW}_{\rm rest}^{\rm H\alpha} \simeq 1960^{+1089}_{-927}~{\rm \AA}$
in $z \sim 8$ star-forming galaxies by stacking the image stamps of 102 Lyman-break galaxy candidates.
The median-staked SED with a blue UV slope ($\alpha_\lambda \simeq -2.4$) and enhanced flux densities at $\lambda_{\rm obs}=3.6$ and $4.5~\mu$m
could be explained by the contribution of a star-forming galaxy with $Z>0.1~\zsun$ and an accreting seed BH.}.

%%%%%%%%
%	Fig. 4     %
%%%%%%%%
\begin{figure*}
\begin{center}
{\includegraphics[width=70mm]{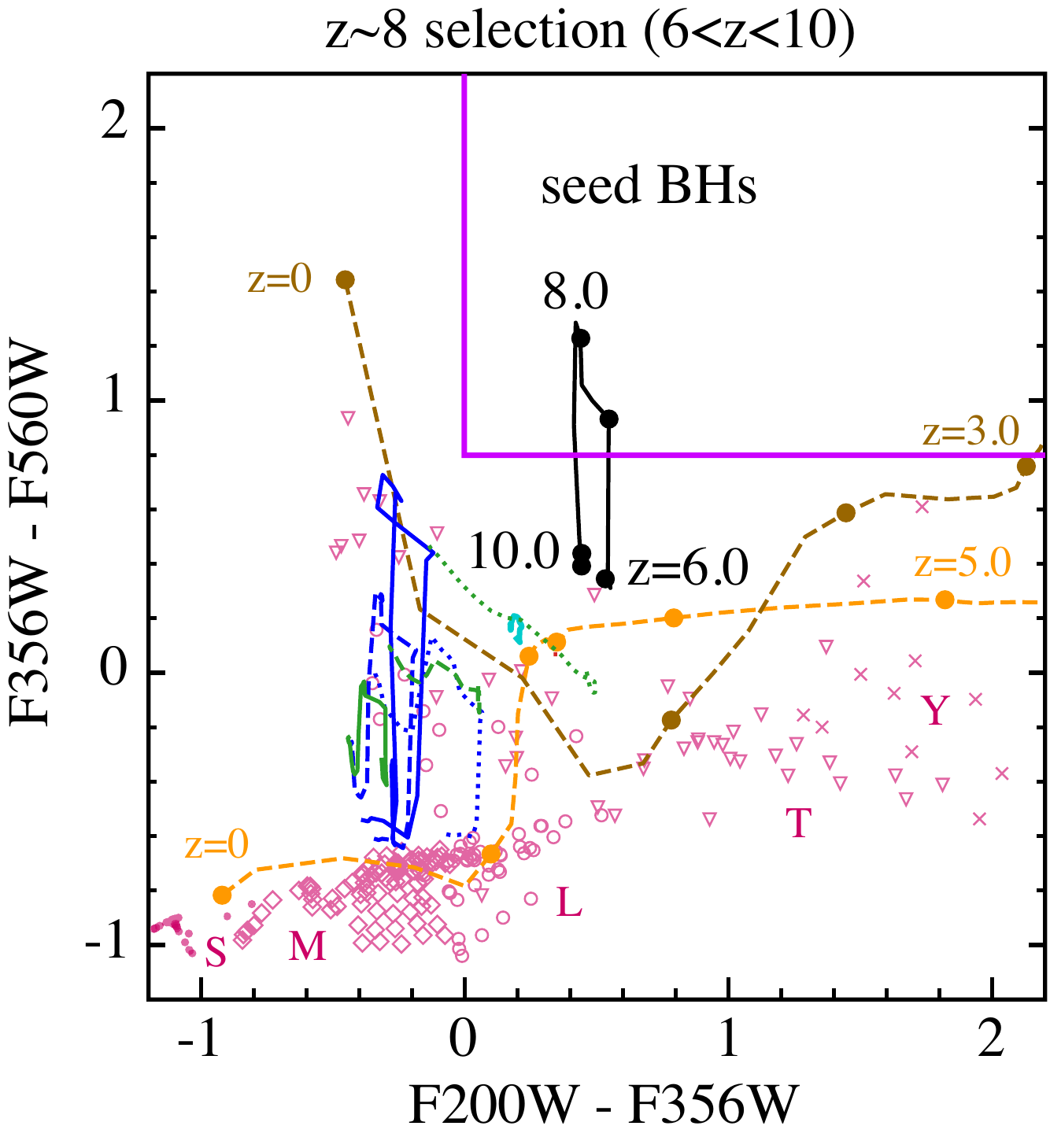}\hspace{15mm}
\includegraphics[width=70mm]{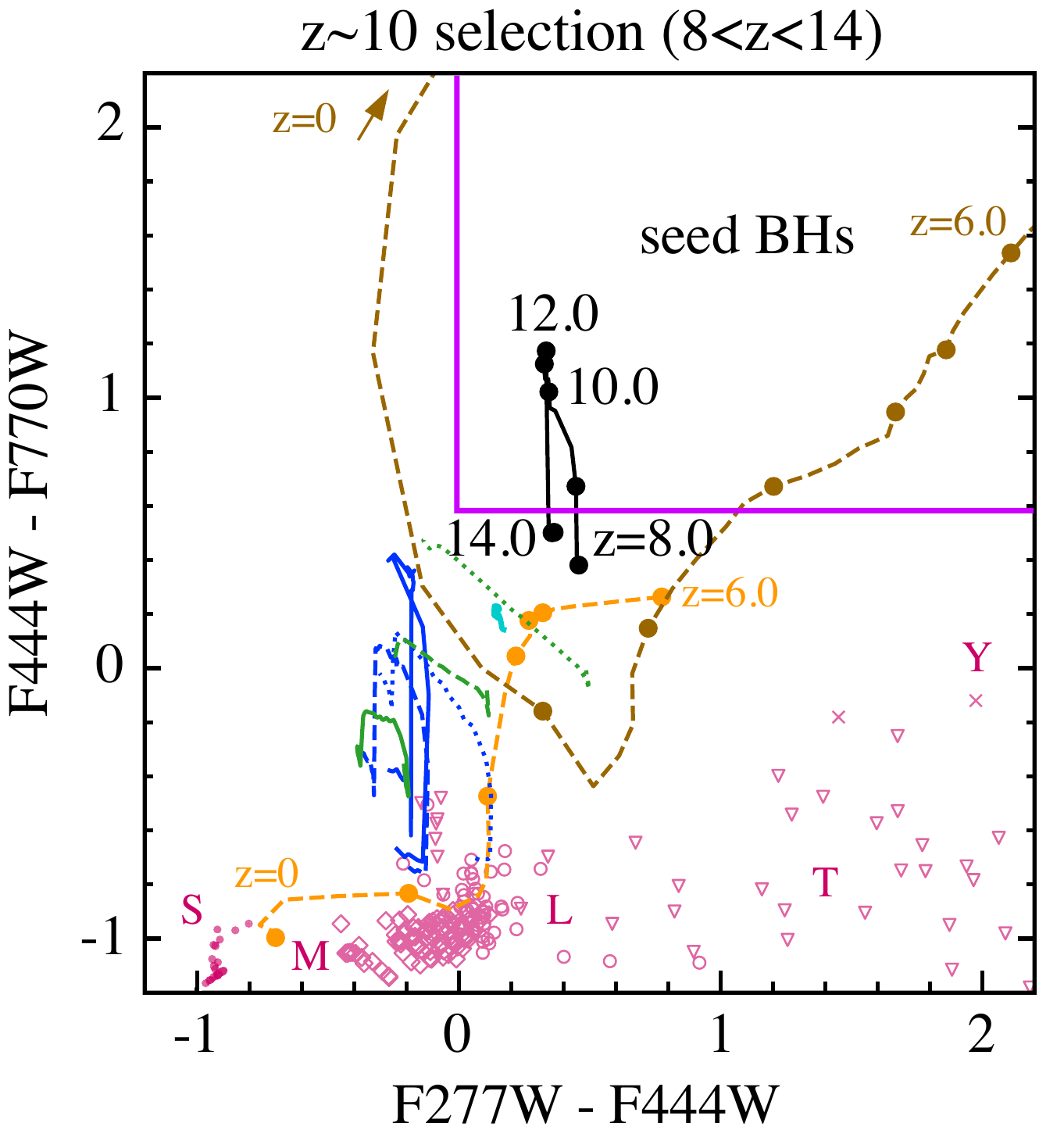}}
\caption{Color-color diagrams of the growing seed BHs at $z\sim8$ (left) and $z\sim10$ (right).
We show the same objects as in Figure~\ref{fig:color} with the same line styles.
Only the colors in limited redshift ranges ($6 \leq z \leq 10$ for the F200W$-$F356W vs F356W$-$F560W plane, 
and $8 \leq z \leq 14$ for the F277W$-$F444W vs F444W$-$F770W plane) are presented.
We also show other galaxy populations:
Balmer-break galaxies (orange) and extremely dusty star-forming galaxies (brown), which are obtained from the galaxy SED models used in \citet{Mawatari_2020}.
For visualization we overplot the colors with filled dots with an interval of $\Delta z=1$.
Galactic brown dwarfs from the BT-Settl model \citep[][solar-metallicity model only]{Allard_2012} are shown with magenta open symbols: M (diamond), L (circle), T (triangle), and Y (cross).
O-to-M stars (magenta dot) are computed from \cite{Kurucz_1993}.
The color-cut conditions given in Eqs.~(\ref{eq:ccz8}) and (\ref{eq:ccz10})
provide robust criteria for photometric selection of rapidly growing BHs  (magenta regions). 
See the main text for more details.}
\label{fig:color2}
\end{center}
\end{figure*}

%%%%%%%%
%	Fig. 5     %
%%%%%%%%
\begin{figure*}
\begin{center}
{\includegraphics[width=70mm]{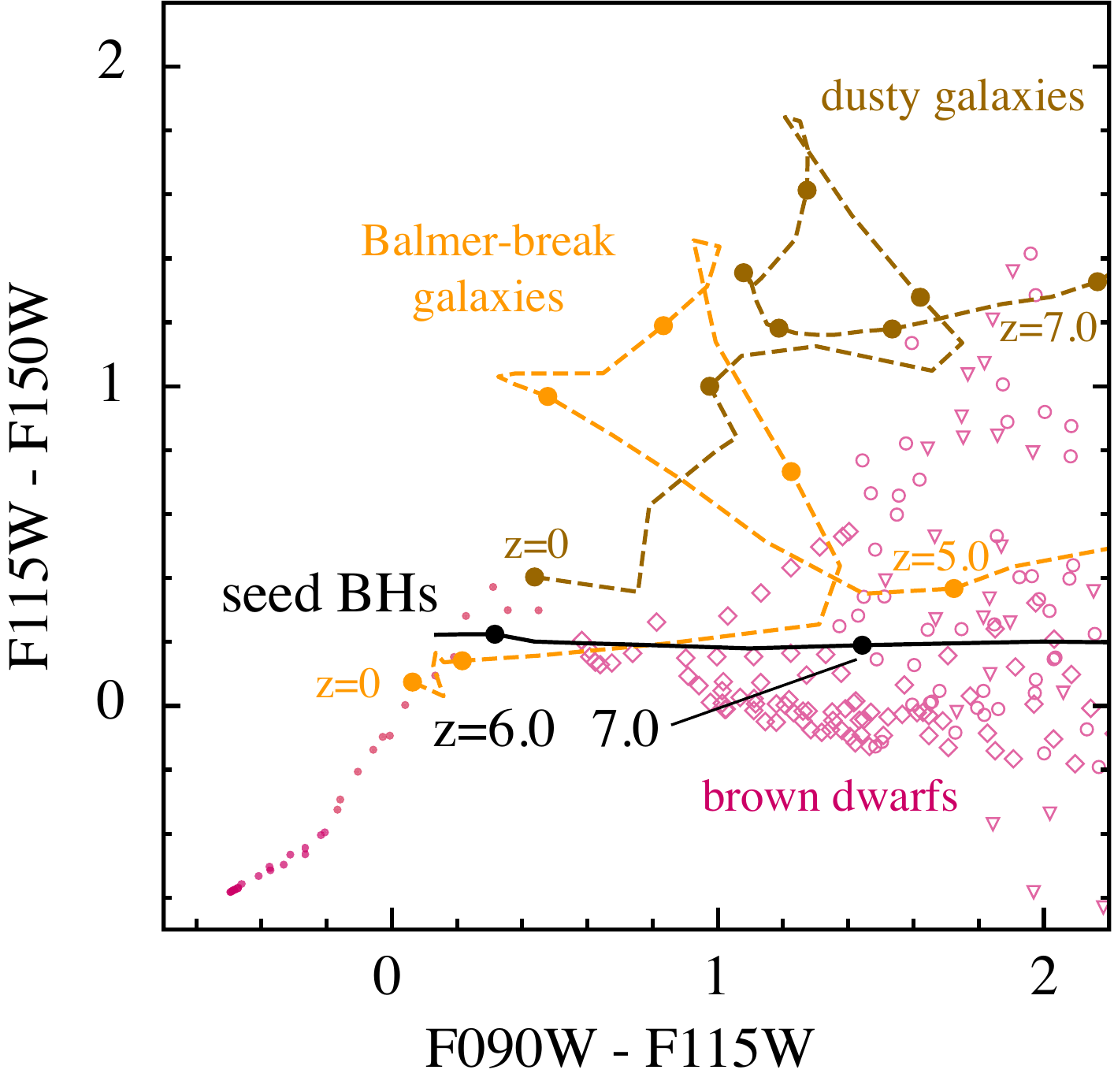}\hspace{15mm}
\includegraphics[width=70mm]{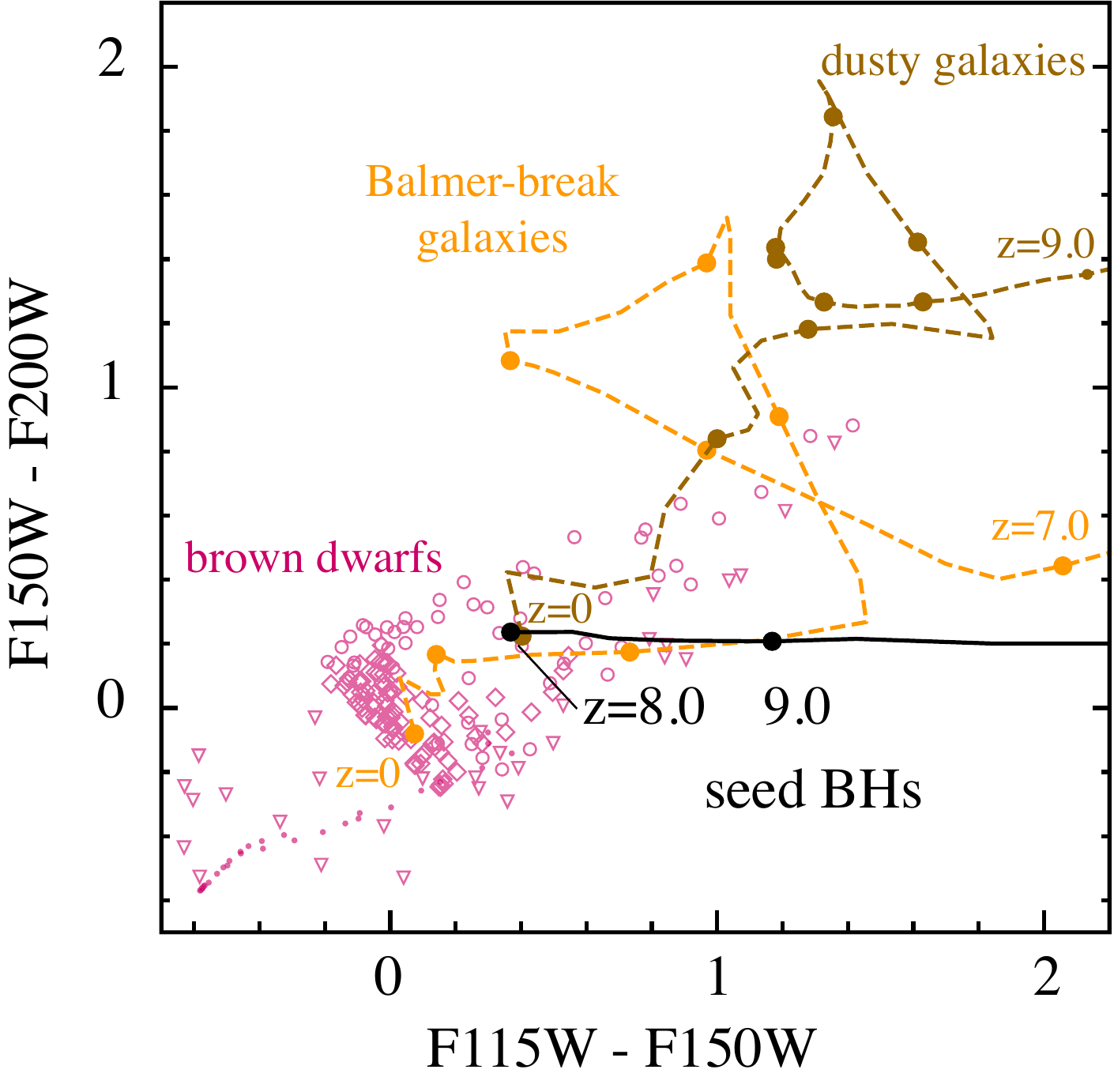}}
\caption{Color-color diagrams for NIRCam dropout selection at $z\sim8$ (left) and $z\sim10$ (right).
The colors of high-$z$ seed BHs at $6<z<9$ (black) are heavily contaminated by Galactic brown dwarfs 
(open symbols), as is also the case for Lyman-break galaxies \citep{Hainline_2020},
but are clearly separated from extremely red galaxies at intermediate redshifts of $z>1$.
}
\label{fig:colorNIR}
\end{center}
\end{figure*}

% color-color plot
Figure~\ref{fig:color2} shows the location of such systems at $z\sim 8$ (left) and $z\sim 10$ (right)
in the color-color space of the broad-band filters of interest.
Rapidly growing seed BHs are characterized by extremely red infrared colors caused by the flat continuum of the BH radiation and strong H$\alpha$ emission.
As demonstrated in Figure~\ref{fig:color}, accreting seed BHs tend to be redder compared to the other three cases 
at the same redshift range and thus can be distinguishable by the following color cuts, for example;
\begin{equation}
z\sim 8:
\begin{cases}
{\rm F200W}-{\rm F356W} >0,\\[3pt]
{\rm F356W}-{\rm F560W} > 0.8,
\end{cases}
\label{eq:ccz8}
\end{equation}
%
%and
%
\begin{equation}
z\sim 10:
\begin{cases}
{\rm F277W}-{\rm F444W} > 0, ~~~~~~~~\\[3pt]
{\rm F444W}-{\rm F770W}>0.6.
\end{cases}
\label{eq:ccz10}
\end{equation}
Those color conditions are denoted by the magenta boundaries in Figure~\ref{fig:color2}.
As discussed in \S\ref{sec:SED}, 10~(20) ks exposures of the NIRCam and MIRI filters can detect 
the accreting seed BHs at $z\sim 8~(10)$ with the significance of S/N$\ga 10~(5)$.
Therefore, within the expected $1\sigma$ photometric error of $< 0.1~(0.2)$ mag,
the conditions above enable the color selection of the seed BHs in the two redshift ranges.

In practice, there still remains possible contamination from nearby and intermediate-redshift astrophysical objects.
Galactic brown dwarfs yield very red colors in the NIRCam filters \citep{Allard_2012} and 
show point-like appearance \citep{Matsuoka_2016}.
Strong Balmer-break galaxies and extremely dusty star-forming galaxies at 
intermediate redshifts also show extremely red colors in near-infrared filters \citep[e.g.,][]{Mawatari_2020}.
In Figure~\ref{fig:color2}, we overlay the colors of these types of red objects.
Overall, most of them appear outside the region given by the color selection cuts
(note that dusty galaxies at $z\gtrsim3$ can be removed by Lyman dropout selection with NIRCam filters; see below).
This clearly shows the importance of detecting H$\alpha$ emission with MIRI to separate high-$z$ seed BHs
from other astrophysical contaminants.

We presume that the growing seed BHs would first be selected as dropout sources, as they are affected by strong IGM
absorption just as well as other high-$z$ galaxies.
Figure~\ref{fig:colorNIR} shows the color-color plots that are proposed for NIRCam selection of $z\sim8$ and $z\sim10$ Lyman-break galaxies in \citet{Hainline_2020}.
The filters in the first colors, F090W$-$F115W ($z\sim8$) and F115W$-$F150W ($z\sim10$), are chosen to straddle the redshifted Lyman break, which are red for the $z>7$ and $z>9$ sources, respectively.
The second colors, F115W$-$F150W ($z\sim8$) and F150W$-$F200W ($z\sim10$), are modestly red ($\sim0.2$), as the third filters cover redward of Ly$\alpha$, which is not affected by IGM absorption.
We find that the growing seed BHs are seriously contaminated by Galactic brown dwarfs in the $z\sim8$ selection \citep[see][]{Hainline_2020}.
This issue is more serious in our case only with the NIRCam colors, as we consider sources that have a continuum redder than that of normal unobscured galaxies.

As shown in Figure~\ref{fig:color2}, the $z\gtrsim3$ dusty galaxies fall relatively close to the seed BH regime and would be potential contaminants.
However, those galaxies are extremely red ($\gtrsim 1$) in all the NIRCam colors we show in Figure~\ref{fig:colorNIR}.
Therefore, we conclude that the low-$z$ galaxies are mostly removed in the dropout selection for $z\sim8$ and $z\sim10$ sources.
Figure~\ref{fig:colorNIR} also shows that the Balmer breaks of low-$z$ unobscured galaxies mimic Lyman breaks and contaminate the dropout selection. 
However, those mature galaxies have a flatter continuum (or bluer colors) redward of the Balmer break (Figure~\ref{fig:color2}).
In summary, contamination of seed BH selection by dusty and Balmer-break galaxies is likely not a serious problem.

Several previous studies have modeled the radiation spectrum of an accreting seed BH and have discussed their colors for JWST observations.
Based on a spherically symmetric, one-dimensional flow structure to produce synthetic spectra,
the predicted SEDs tend to show red colors in the NIRCam bands \citep{Pacucci_2015, Natarajan_2017, Valiante_2018}.
\cite{Barrow_2018} discussed the detectability of accreting seed BHs using their three-dimensional cosmological simulations
and proposed a color-cut condition.
Previous studies claim that most color cuts with NIRCam filters can 
distinguish accreting seed BHs distinguished from young star-forming 
galaxies.  However, in the absence of additional constraints from X-ray counterparts \citep{Natarajan_2017},
their color selection (e.g., F090W$-$F200W versus F200W$-$F444W) are seriously contaminated by Galactic brown 
dwarfs and extremely red galaxies at intermediate redshifts.

\section{Discussion and Conclusions}

Photometrically detected objects that show colors of high-$z$ seed BHs are ideal targets for follow-up spectroscopic observations.
In Table~\ref{tab:EW}, we show the signal-to-noise ratio (S/N per pixel) of each line of the SED ($z=8$), assuming a line width of $300~\kms$ and
a NIRSpec exposure time of 30 ks, for both the low-resolution (PRISM; $R \sim 100$) 
and medium-resolution ($R \sim 1,000$) modes. 
In both spectroscopic modes, all the emission lines except \ion{He}{2} can be significantly detected with S/N $\ga 4$, allowing us
to diagnose the gas properties 
in the circumnuclear region at $<10~\pc$ of the protogalaxy and to infer the mass accretion rate 
through the disk at $0.1-1~\pc$.
Detailed modeling of these relevant lines and their physical interpretation will be investigated in forthcoming work.

In addition to narrow lines, some broad components (especially hydrogen lines, owing to the lack of metallicities) should be detectable, 
which would further enhance the observed line fluxes.
The broad component of H$\alpha$ or H$\beta$, if detected, would be useful for single-epoch BH mass measurement 
\citep[e.g.,][]{Greene_Ho_2005,Ho_Kim_2015}.
Combining the BH mass with the inferred accretion rate in the disk based on \ion{O}{1} and \ion{C}{2}]
emission lines, one can also estimate the Eddington ratio of the accretion rate of the growing seed BHs
and thus explore the relation between the luminosity and accretion rate of such extreme objects.

In conclusion, the upcoming JWST observations, with the aid of physically-motivated selection criteria,
will be able to improve the detection efficiency of fast-growing seed BHs.
The discovery of such unique objects will prove that seed BHs had experienced super-Eddington growth early on \citep{Inayoshi_2022,Hu_2022b}, 
offering a path toward explaining the 
population of overmassive BHs seen in high-$z$ quasars \citep[e.g.,][]{Izumi_2021}.
Observations of such BHs and their host galaxies will be a milestone in revealing how coevolution between BHs and host galaxies has 
been established at cosmic dawn \citep{Habouzit_2022}.
The arrival of the JWST will usher in a new age of discovery. Our calculations show that JWST can be used to excavate one piece of the
fundamental missing links for the structure formation in the early universe.

\acknowledgments
We greatly thank Xuheng Ding, Takuya Hashimoto, Tohru Nagao, and John Silverman for constructive discussions. 
We also thank Ken Mawatari for sharing the galaxy SED model templates.
We acknowledge support from the National Natural Science Foundation of China (12073003, 12003003, 11721303, 11991052, 11950410493), 
and the China Manned Space Project with NO. CMS-CSST-2021-A04 and CMS-CSST-2021-A06. 
Y. S. and A. K. I. acknowledge support from NAOJ ALMA Scientific Research Grant Code 2020-16B.
The numerical simulations were performed with the Cray XC50 at the Center for Computational Astrophysics (CfCA) of the National Astronomical 
Observatory of Japan and with the High-performance Computing Platform of Peking University.

\bibliography{ref}{}
\bibliographystyle{aasjournal}
%% This command is needed to show the entire author+affiliation list when
%% the collaboration and author truncation commands are used.  It has to
%% go at the end of the manuscript.
%\allauthors

%% Include this line if you are using the \added, \replaced, \deleted
%% commands to see a summary list of all changes at the end of the article.
%\listofchanges

\end{document}